%% file: ctp.v.2.9.tex
\documentclass[prl,twocolumn,floatfix,superscriptaddress,showpacs]{revtex4}
\usepackage{graphicx,dcolumn,bm}
\providecommand{\EPJC}{Eur. Phys. J. C }

\providecommand{\NIMA}{Nucl. Instr. Meth. A }

\providecommand{\NPB}{Nucl. Phys. B }

\providecommand{\PLB}{Phys. Lett. B }

\providecommand{\PRL}{Phys. Rev. Lett. }
\providecommand{\PRC}{Phys. Rev. C }
\providecommand{\PRD}{Phys. Rev. D }
\providecommand{\RMP}{Rev. Mod. Phys. }

\providecommand{\qqb}{$q\bar{q}$~}
\begin{document}

\title{ The  ${\bm Q^{2}}$-Dependence of Nuclear Transparency for 
Exclusive ${\bm \rho^0} $ Production \\
 }
\include{authors}
 \date{\today.  To be submitted to Phys. Rev. Lett.}


\begin{abstract}
Exclusive coherent and incoherent electroproduction of the $\rho^0$
meson from $^1$H and $^{14}$N targets has been studied at the HERMES
experiment as a function of coherence length ($l_c$), corresponding to 
the lifetime of hadronic fluctuations of the virtual photon, and squared
four-momentum of the virtual photon ($-Q^2$). The ratio of $^{14}$N to
$^1$H cross sections per nucleon, known as nuclear transparency, was found 
to increase (decrease) with increasing coherence length for coherent
(incoherent) $\rho^0$ electroproduction. For fixed coherence length,
a rise of nuclear transparency with $Q^2$ is observed for both
coherent and incoherent $\rho^0$ production, which is in agreement
with theoretical calculations of color transparency.
\end{abstract}

\maketitle

One of the fundamental predictions of QCD is the existence of a
phenomenon called color transparency (CT), whose characteristic
feature is that, at sufficiently high squared four-momentum
transfer $(-Q^2)$ to a hadron, the initial state  and final state 
interactions of that hadron traversing a nuclear medium 
vanish \cite{Kop81,Ber81,Bro88,Nik92,Kop93,Bro94,Fra94}. The idea is that 
the dominant amplitudes for exclusive reactions at high $Q^2$ involve 
hadrons of reduced transverse size, and that these  small color-singlet 
objects or small size configurations (SSC) have reduced interactions with
hadrons in the surrounding nuclear medium. Moreover, it is assumed 
that these SSC remain small long enough to traverse the nucleus.

Several experiments in search of CT \cite{Car88,Mak94,One95,Abb98}
have been carried out over the last fifteen years.
Although none of these experiments is in conflict with CT only a few
have shown evidence for it. The strongest evidence for CT 
so far comes from Fermilab experiment E791 on the $A$-dependence 
of coherent diffractive dissociation of 500 GeV/c pions into 
di-jets \cite{Ait01}. This result shows a platinum to carbon 
cross section about ten times larger than expected if soft processes 
would dominate, which is qualitatively consistent with 
theoretical calculations of CT effects \cite{Fra93,Fra00}. 
Also experiment E665 on exclusive incoherent $\rho^0$ 
muoproduction from nuclei \cite{Ada95} gives an indication of 
CT. However, that signal is of indecisive statistical significance. 
In this letter we report new  evidence of CT in 
exclusive coherent and incoherent $\rho^0$ electroproduction 
using a novel analysis technique. 

When searching for CT, a commonly used observable
is the nuclear transparency $T=\sigma_A/(A\sigma_p)$, which is the 
ratio of the nuclear cross section per nucleon to that on the 
proton. For diffractive incoherent reactions on a nuclear
target, where the nucleus is excited or breaks up,
CT predicts that as $Q^2$ becomes large, $T$ approaches
unity, independent of $A$. For coherent reactions, where the
interaction leaves the whole nucleus intact in its ground
state, the same observable is used in spite of the fact that
it can no longer be associated directly with the probability
of escape of the hadron from the nucleus. This complication
arises because of the strong influence of the nuclear form
factor, which is sensitive to the kinematics of the 
reaction.

In order to study CT for exclusive electroproduction of $\rho^0$
mesons, one has to select a sample of $\rho^0$ mesons produced 
by photons with large $Q^2$. In these processes, the hadronic 
structure of a high energy virtual photon \cite{Bau78} in the form 
of a \qqb pair has a transverse size $r_\perp \sim 1/Q$~\cite{Bro88}. 
The \qqb fluctuation of the virtual photon can propagate over a distance 
$l_c$ known as the coherence length. It is given~\cite{Bau78,Nik92} 
by $l_c=\frac{2\nu}{Q^{2}+M_{q\bar{q}}^{2}}$, where $\nu$ is the virtual 
photon energy and $M_{q\bar{q}}$ is the invariant mass of the \qqb pair. 
This SSC can propagate through the nuclear medium with little interaction.
After the \qqb pair is put on-shell, it will evolve to a normal-size
$\rho^0$ meson over a distance  $l_f$ called the formation length.
It is a governing scale for the CT effect and is given~\cite{Bro88} 
by $l_f =\frac{2\nu}{m_{V^\prime}^2 - m_V^2}$, where $m_V$ is the mass of 
the $\rho^0$ meson in the ground state and $m_{V^\prime}$ the mass of its
first radial excitation.

The phenomena determining nuclear transparency form an intricate mixture 
of coherence and formation length effects. They have a different 
appearance for coherent and incoherent $\rho^0$ production.
For incoherent  production, the probability for the \qqb pair 
to interact with the nuclear medium increases with  $l_c$ until
$l_c$ exceeds the nuclear size~\cite{Huef96,clincoh}.  
For  values of the $l_c$ smaller than the nucleus, this coherence length
effect~\cite{clincoh}
can mimic the $Q^2$-dependence of the nuclear transparency
predicted by CT. For coherent $\rho^0$ production, in contrast,
the nuclear form factor suppresses the apparent
nuclear transparency. Small $l_c$
corresponds to a large longitudinal momentum transfer ($q_{c}
\sim 1/l_c$), where the form factor is small. Hence, $T$ decreases
with $Q^2$ in coherent production. This behaviour cannot mimic CT, 
but also in this case, the coherence length effects can significantly
modify the $Q^2$-dependence, thus obscuring the clean observation
of a CT effect. In order to disentangle the effects of coherence length 
from those of CT, it is important to study the variation of $T$ with
$Q^2$, while keeping $l_c$ fixed~\cite{Kop02}. In this way, a change of 
$T$ with $Q^2$ can be associated with the onset of CT.

A rigorous quantum mechanical description of the SSC evolution,
based on the light-cone Green function formalism \cite{Kop02},
naturally incorporates the effects of both coherence length
and CT. In this formalism it is shown
that the signature of CT is a positive slope of the
$Q^2$-dependence of  nuclear transparency at fixed coherence length
for both coherent and incoherent $\rho^0$ production.
We have sought this signature. 


The data were obtained during the 1996-1997 running periods of the
HERMES experiment in the 27.5 GeV HERA positron storage ring at
DESY. Stored beam currents ranged from 5 to 40 mA. Integrated
luminosities of 108 and 50 pb$^{-1}$ were collected on $^1$H and
$^{14}$N internal gas targets, respectively. The corresponding
average target thicknesses were $10^{14}$ and $10^{15}$
nucleons/cm$^2$. The thicknesses were additionally varied by factors of
0.5 to 10, depending on how much the HERMES internal
target was allowed to limit the HERA beam lifetime. The scattered
$e^+$ and the $\pi^+ \pi^-$ pair from the $\rho^0$ decay were
detected in the HERMES spectrometer \cite{hermes_spec}.


The $\rho^0$ production sample was extracted from events with
exactly three tracks: a scattered positron and two oppositely charged
hadrons, as  described in detail in Ref.~\cite{clincoh}.
Evaluated for each event were the Bjorken scaling variable 
$x = Q^{2} / 2 m_{p} \nu$, with $m_p$ the mass of the proton, 
the squared four-momentum
transfer to the target $t^{\prime} =t - t_{0}$, with $t_{0}$ its
minimum value, and the photon-nucleon invariant mass squared
$W^{2} = m_{p}^{2} + 2 m_{p} \nu - Q^{2}$. The
kinematic coverage in $\nu$, $x$ and $W$ is 
$5<\nu<24$ GeV, $0.01<x<0.35$ and $3<W<6.5$
GeV, with mean values of 13.3 GeV, 0.07 and 4.9 GeV, respectively.
 
The exclusive
$\rho^0$ production signal was extracted in the kinematic region 
$-2 < \Delta E < 0.6$ GeV and  $0.6 < M_{\pi \pi} < 1$ GeV, where
$\Delta E=\nu-E_{\rho}+\frac{t}{2m_{p}}$ is the exclusivity 
variable~\cite{clincoh,rhoxsect} with $E_{\rho}$ the energy of the 
produced $\rho^0$ meson, and $ M_{\pi \pi}$ the invariant mass of 
the detected 
hadron pair, assuming that they were pions. In the analysis of 
nuclear transparency for coherent production, the
exclusive $\rho^0$ mesons have been selected with
$|t^\prime| < 0.045$ GeV$^{2}$ for nitrogen and $|t^\prime| < 0.4$
GeV$^{2}$ for hydrogen, while in the analysis for incoherent production 
the $t^\prime$ restriction was $0.09<|t^\prime| < 0.4$ GeV$^{2}$ for both data
samples. The resolution of $\Delta E$ is about 0.25 GeV~\cite{rhoxsect}, 
and the $t^\prime$ resolution is about 0.008 GeV$^2$.
It has been shown \cite{clincoh} that the incoherent
$t^{\prime}$ slope parameter $b_p$  for various nuclei is
consistent with the  hydrogen value $b_p = (7.08 \pm 0.3)$
GeV$^{-2}$, and that the observed $Q^2$-dependence of $b_p$ agrees
well with other existing data \cite{tytgat}. The coherent slope parameter
on nitrogen, $b_{^{14}\rm N} = (57.2 \pm 3.3)$ GeV$^{-2}$, is in
agreement with the values predicted by the relationship $b_{A}
\approx R_{A}^{2}/3$ \cite{Alv70}, where $R_{A}$ is the nuclear
radius.


The background under the exclusive $\rho^0$ peak has been
described earlier \cite{clincoh}. It is mainly caused by hadrons
from semi-inclusive deep inelastic scattering (DIS). Part
of this background is removed by excluding the region
$|t^{\prime}| > 0.4 $ GeV$^2$, where the background dominates the
$\rho^0$ yield. The remaining background $(6\pm 3\%)$ is estimated
from the number of events measured at $|t^{\prime}| < 0.4 $ GeV$^2$ 
and $\Delta E > 3$ GeV, and subtracted \cite{clincoh}. The
double-diffractive contribution to the incoherent $\rho^0$
production cross section is found to be $(4 \pm 2\%)$
\cite{rhoxsect}, for which the data were corrected. For coherent
$\rho^0$ production the contamination due to double-diffractive
dissociation is found to be negligible.


For incoherent $\rho^0$ production the nuclear transparency 
has been
evaluated as $T_{inc} = \sigma^A_{inc} / (A \sigma_{p}) = 
N_{inc}^A {\mathcal L}_{p} / (A N_{p} {\mathcal L}_{A})$ \cite{clincoh}, 
where $p$ refers to $^1$H, $A$ to $^{14}$N, $N_{inc}^A$ is the 
number of incoherent events, $ N_{p}$ the number of events on  $^1$H,
and ${\mathcal L}_{A,p}$ the effective luminosity of the
nitrogen or hydrogen samples, corrected for detector and
reconstruction inefficiencies.  In addition, a ``Pauli blocking''
correction has been applied, which
accounts for the absence of incoherent $\rho^0$ production on 
nitrogen at momentum transfers $|t^\prime|$ smaller than the nuclear 
Fermi momentum \cite{Tref69}.

For coherent $\rho^0$ production the quantity 
$T_{c} = \sigma^A_{c} / (A \sigma_{p})$ is evaluated. 
Some additional correction
factors have been applied to $\sigma^A_{c}$ to extract  $T_{c}$
because  different $t^\prime$ requirements have been applied to
the nitrogen and the hydrogen data. These include the ratio of
the different acceptance correction factors caused by the
different $t^\prime$ regions selected, which has been obtained
from Monte Carlo simulations of exclusive $\rho^0$ production in
a $4\pi$ geometry and in the HERMES acceptance \cite{rhoxsect};
the radiative correction factors \cite{radcor}, which were calculated
separately for nitrogen and hydrogen  for each
$l_c$ bin; and the contamination of incoherent background in
the coherent sample. No $Q^2$-dependence has been observed for the ratios of  
any of the correction factors. 
 
\begin{figure}[!htbp]
 \vspace*{-0.5cm}
  \begin{center}
   \includegraphics[height=75mm]{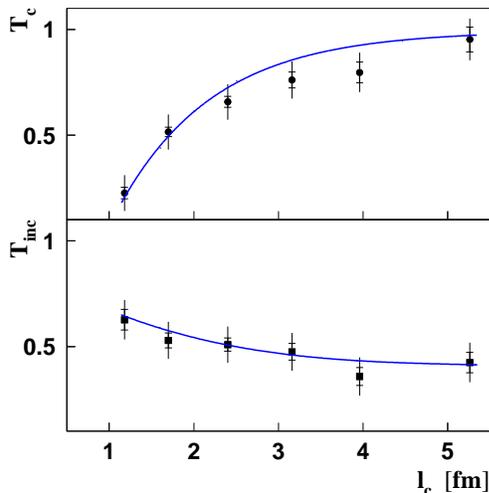}
  \end{center}
   \vspace*{-1.cm}
  \caption{Nuclear transparency as a function of coherence length for
  coherent (top panel) and incoherent (bottom panel) $\rho^0$
  production on nitrogen, compared to predictions with CT effects 
  included (curves)~\cite{Kop02}. The inner error bars include only statistical
  uncertainties, while the outer  error bars present the statistical 
  and  systematic uncertainties added in quadrature.}
\label{fig:fig1}
\end{figure}

The nuclear transparencies for coherent and incoherent $\rho^0$ 
production are presented in Fig.~\ref{fig:fig1}. The data for incoherent 
$\rho^0$ production supersede the previously published data
\cite{clincoh}, as the present analysis includes ``Pauli blocking''
corrections and the same requirement on  $t^\prime$ over the entire 
coherence length region.
The data decrease with increasing  $l_c$, as expected from the effects of 
initial state interactions.

The nuclear transparency for coherent $\rho^0$ production increases
with coherence length as expected from the effects of the nuclear 
form factor~\cite{Kop02}. 
Good agreement is found between the measured nuclear transparencies,
integrated over the available $Q^2$ region, and calculations
including both the coherence length and CT effects  ~\cite{Kop02}
evaluated for each $l_c$ bin at their mean experimental $l_c$ and $Q^2$
values, given by the curves in Fig.~\ref{fig:fig1}.
The effect of the nuclear form factor on $T_c$ is included in the 
calculations. Moreover, when extracting the $Q^2$-dependence of $T_c$
for separate $l_c$ bins, the nuclear form factor effect does not contribute
as it depends on $t$ rather than $Q^2$~\cite{Kop02}. 
Hence, the slope with $Q^2$ is unaffected if $l_c$ (and hence to a very good
approximation $t$) is fixed. 

The systematic uncertainties are separated into 
$Q^2$- and $l_c$-dependent and kinematics-independent contributions. The
ratio of the integrated luminosities represents the largest source of
kinematics-independent uncertainties. An additional contribution comes
from double-diffractive dissociation. The total estimated systematic 
uncertainty from all
normalization factors is $11\%$. The kinematics-dependent
systematic uncertainties have been studied as a function of $l_c$
and $Q^2$ on a bin-by-bin basis. The main contributions come from
DIS background subtraction, from acceptance corrections, the
efficiency of the $\Delta E$ exclusivity cut, the 
corrections due to ``Pauli blocking'', and the application of
radiative corrections. None of the kinematics-dependent systematic
uncertainties cancel in the coherent nuclear transparency
because of the different $t^\prime$ cuts that
are applied, and they increase at small and at large coherence
length values. At small $l_c$, and correspondingly large $Q^2$,
the uncertainties in the coherent to incoherent separation via the 
$t^\prime$ slope parameters $b_p$ and $b_{^{14}\rm N}$, 
and the background subtraction dominate. At
large $l_c$, the uncertainty in the acceptance correction factor
becomes large. Thus, the contribution of the kinematics-dependent
systematic uncertainty varies between $8\%$ and $14\%$. 
This results in a combined systematic uncertainty of $14\%$ to $18 \%$
for the nuclear transparency measurements 
presented in Fig.~\ref{fig:fig1}. 

\begin{figure}[!htbp]
 \vspace*{-0.5cm}
  \begin{center}
   \includegraphics[height=75mm]{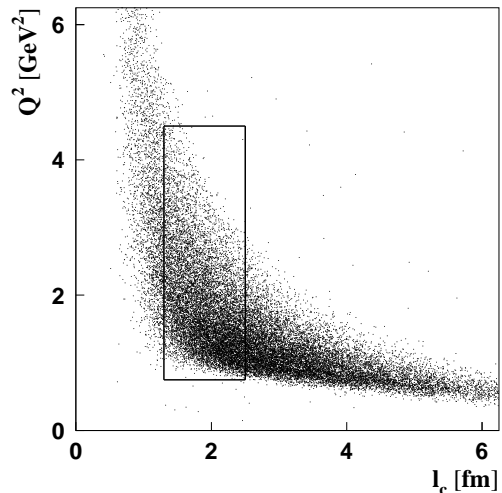}
  \end{center}
 \vspace*{-1.cm}
  \caption{Distribution of $Q^2$ versus coherence length
  for exclusive $\rho^0$ production on hydrogen and nitrogen.
  The region surrounded by the rectangle represents the subset that was 
  used for the two-dimensional analysis of the nuclear transparency.}
 \label{fig:fig2}
\end{figure}

A two-dimensional analysis of the nuclear transparency as
a function of coherence length and $Q^2$ has been performed, which
represents a new approach in the
search for CT. It is constrained by the phase space boundaries
displayed in Fig.~\ref{fig:fig2}. Since the combination of
statistical significance and $Q^2$ coverage is largest near $l_{c}
\simeq 2.0$ fm, the region $1.3 < l_c < 2.5$ fm has been chosen
for this two-dimensional analysis. To deconvolute the CT and
coherence length effects, coherence length bins of 0.1 fm were used.
These finite bins introduce an additional  systematic uncertainty 
in the $Q^2$-slope of 0.008 and 0.004 GeV$^{-2}$ for 
coherent and incoherent $\rho^0$ production, respectively. In order to 
extract the $Q^2$-dependence, each $l_c$ bin was independently split 
into 3 or 4 $Q^2$ bins. 
  
\begin{figure}[!htbp]
 \vspace*{-0.5cm}
  \begin{center}
   \includegraphics[height=75mm]{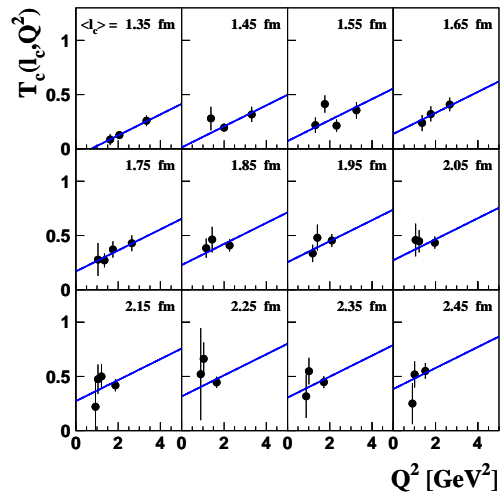}
 \end{center}
  \vspace*{-1.cm}
 \caption{Nuclear transparency as a function of $Q^2$ in specific
  coherence length bins (as indicated in each panel) for coherent
  $\rho^0$ production on nitrogen. The straight line is the result of
  the common fit of the $Q^2$-dependence. The error bars include
  only statistical uncertainties.}
\label{fig:fig3}
\end{figure}

\begin{figure}[!htbp]
 \vspace*{-0.5cm}
 \begin{center}
   \includegraphics[height=75mm]{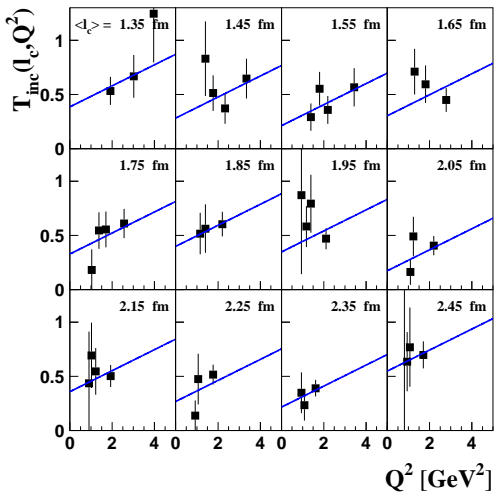}
  \end{center}
  \vspace*{-1.cm}
  \caption{As for Fig. 3, except here for incoherent production.}
\label{fig:fig4}
\end{figure}

The nuclear transparency was extracted in each $(l_c, Q^2)$
bin, and is shown in Figs.~\ref{fig:fig3} and~\ref{fig:fig4}
for twelve $l_c$ bins each for coherent and incoherent $\rho^0$
production. The low statistics in each $(l_{c},Q^2)$
bin makes it difficult to fit the slope of the
$Q^2$-dependence for each coherence length bin separately.
Instead, the data have been fitted with a common $Q^2$-slope $(P_1)$,
which has been extracted assuming 
$T_{c(inc)} = \sigma^{^{14}N}_{c(inc)}(l_c,Q^{2}) / A \sigma_{p} =
P_0 + P_1\cdot Q^{2}$, 
letting $P_0$ vary independently in each $l_c$ bin and keeping $P_1$ 
as common free parameter. The results are displayed as the lines in
Figs.~\ref{fig:fig3}  and ~\ref{fig:fig4}. 
In both cases the reduced-$\chi^2$ values are close to unity.
The common slope parameter of the $Q^2$-dependence, $P_1$, represents the
signature of the CT effect averaged over the coherence length
range. This procedure was performed separately for the coherent
and incoherent data. The results of these fits are compared to 
theoretical calculations \cite{Kop02} in Table~\ref{tab:sloperes}.
If the results are combined, the common value and the total uncertainty 
for the slope of the
$Q^2$-dependence of exclusive $\rho^0$ production is ($0.074 \pm
0.021$) GeV$^{-2}$. This is in agreement with the combined theoretical
prediction of about 0.058  GeV$^{-2}$.

In summary, the transparency of the $^{14}$N nucleus to exclusive
coherent and incoherent $\rho^0$ electroproduction was measured by
the HERMES collaboration as a function of both  $Q^2$ and 
the coherence length of \qqb fluctuations of the virtual photon. 
Positive slopes of the $Q^2$-dependence of the nuclear
transparency have been observed on nitrogen for $l_c = 1.3 - 2.5$
fm and $Q^{2} = 0.9 - 4$ GeV$^2$ for exclusive coherent and
incoherent $\rho^0$ production. Those values are in agreement with
theoretical calculations \cite{Kop02}, wherein a positive
slope of the $Q^2$-dependence of the nuclear transparency is
a signature of color transparency. 
This result not only adds further evidence for the existence
of the color transparency phenomenon, but it also elucidates the complex
interplay of various effects on the production of exclusive $\rho^0$ 
mesons at modestly high energies. 

\begin{acknowledgments}
We thank  B.Z. Kopeliovich and J. Nemchik for stimulating discussions
and their calculations, and  S. Brodsky, L. Frankfurt, and M. Strikman
for many fruitful discussions. We gratefully acknowledge the DESY
management for its support and the staff at DESY and the collaborating
institutions for their significant effort, and our funding agencies for
financial support.
\end{acknowledgments}

\begin{table}[!htbp]
\caption{
Fitted slope parameters of the  $Q^2$-dependence of the nuclear
transparency on nitrogen with statistical and systematic uncertainties
given separately.
The results are compared to theoretical predictions  \cite{Kop02}. }
\label{tab:sloperes}
 \begin{ruledtabular}
\begin{tabular}{lll}
Data sample & Measured $Q^2$ slope  &  Prediction  \\
            & (GeV$^{-2}$) &  (GeV$^{-2}$) \\
 \hline
 coherent   & $ 0.070 \pm 0.021 \pm 0.012$ &  0.060 \\
 incoherent & $ 0.089 \pm 0.046 \pm 0.008$ &  0.048 \\
 \end{tabular}
 \end{ruledtabular}
\end{table}

\end{document}

%% file: authors.tex

\def\groupalberta{\affiliation{Department of Physics, University of Alberta, Edmonton, Alberta T6G 2J1, Canada}}
\def\groupargonne{\affiliation{Physics Division, Argonne National Laboratory, Argonne, Illinois 60439-4843, USA}}
\def\groupbari{\affiliation{Istituto Nazionale di Fisica Nucleare, Sezione di Bari, 70124 Bari, Italy}}
\def\groupcolorado{\affiliation{Nuclear Physics Laboratory, University of Colorado, Boulder, Colorado 80309-0446, USA}}
\def\groupdesy{\affiliation{DESY, Deutsches Elektronen-Synchrotron, 22603 Hamburg, Germany}}
\def\groupzeuthen{\affiliation{DESY Zeuthen, 15738 Zeuthen, Germany}}
\def\groupdubna{\affiliation{Joint Institute for Nuclear Research, 141980 Dubna, Russia}}
\def\grouperlangen{\affiliation{Physikalisches Institut, Universit\"at Erlangen-N\"urnberg, 91058 Erlangen, Germany}}
\def\groupferrara{\affiliation{Istituto Nazionale di Fisica Nucleare, Sezione di Ferrara and Dipartimento di Fisica, Universit\`a di Ferrara, 44100 Ferrara, Italy}}
\def\groupfrascati{\affiliation{Istituto Nazionale di Fisica Nucleare, Laboratori Nazionali di Frascati, 00044 Frascati, Italy}}
\def\groupfreiburg{\affiliation{Fakult\"at f\"ur Physik, Universit\"at Freiburg, 79104 Freiburg, Germany}}
\def\groupgent{\affiliation{Department of Subatomic and Radiation Physics, University of Gent, 9000 Gent, Belgium}}
\def\groupgiessen{\affiliation{Physikalisches Institut, Universit\"at Gie{\ss}en, 35392 Gie{\ss}en, Germany}}
\def\groupglasgow{\affiliation{Department of Physics and Astronomy, University of Glasgow, Glasgow G128 QQ, United Kingdom}}
\def\groupillinois{\affiliation{Department of Physics, University of Illinois, Urbana, Illinois 61801, USA}}
\def\groupliverpool{\affiliation{Physics Department, University of Liverpool, Liverpool L69 7ZE, United Kingdom}}
\def\groupwisconsin{\affiliation{Department of Physics, University of Wisconsin-Madison, Madison, Wisconsin 53706, USA}}
\def\groupmit{\affiliation{Laboratory for Nuclear Science, Massachusetts Institute of Technology, Cambridge, Massachusetts 02139, USA}}
\def\groupmichigan{\affiliation{Randall Laboratory of Physics, University of Michigan, Ann Arbor, Michigan 48109-1120, USA }}
\def\groupmoscow{\affiliation{Lebedev Physical Institute, 117924 Moscow, Russia}}
\def\groupmunich{\affiliation{Sektion Physik, Universit\"at M\"unchen, 85748 Garching, Germany}}
\def\groupnikhef{\affiliation{Nationaal Instituut voor Kernfysica en Hoge-Energiefysica (NIKHEF), 1009 DB Amsterdam, The Netherlands}}
\def\groupstpetersburg{\affiliation{Petersburg Nuclear Physics Institute, St. Petersburg, Gatchina, 188350 Russia}}
\def\groupprotvino{\affiliation{Institute for High Energy Physics, Protvino, Moscow oblast, 142284 Russia}}
\def\groupregensburg{\affiliation{Institut f\"ur Theoretische Physik, Universit\"at Regensburg, 93040 Regensburg, Germany}}
\def\grouprome{\affiliation{Istituto Nazionale di Fisica Nucleare, Sezione Roma 1, Gruppo Sanit\`a and Physics Laboratory, Istituto Superiore di Sanit\`a, 00161 Roma, Italy}}
\def\groupsimonfraser{\affiliation{Department of Physics, Simon Fraser University, Burnaby, British Columbia V5A 1S6, Canada}}
\def\grouptriumf{\affiliation{TRIUMF, Vancouver, British Columbia V6T 2A3, Canada}}
\def\grouptokyo{\affiliation{Department of Physics, Tokyo Institute of Technology, Tokyo 152, Japan}}
\def\groupamsterdam{\affiliation{Department of Physics and Astronomy, Vrije Universiteit, 1081 HV Amsterdam, The Netherlands}}
\def\groupwarsaw{\affiliation{Andrzej Soltan Institute for Nuclear Studies, 00-689 Warsaw, Poland}}
\def\groupyerevan{\affiliation{Yerevan Physics Institute, 375036 Yerevan, Armenia}}


\groupalberta
\groupargonne
\groupbari
\groupcolorado
\groupdesy
\groupzeuthen
\groupdubna
\grouperlangen
\groupferrara
\groupfrascati
\groupfreiburg
\groupgent
\groupgiessen
\groupglasgow
\groupillinois
\groupliverpool
\groupwisconsin
\groupmit
\groupmichigan
\groupmoscow
\groupmunich
\groupnikhef
\groupstpetersburg
\groupprotvino
\groupregensburg
\grouprome
\groupsimonfraser
\grouptriumf
\grouptokyo
\groupamsterdam
\groupwarsaw
\groupyerevan


\author{A.~Airapetian}  \groupyerevan
\author{N.~Akopov}  \groupyerevan
\author{Z.~Akopov}  \groupyerevan
\author{M.~Amarian}  \grouprome \groupyerevan
\author{V.V.~Ammosov}  \groupprotvino
\author{A.~Andrus}  \groupillinois
\author{E.C.~Aschenauer}  \groupzeuthen
\author{W.~Augustyniak}  \groupwarsaw
\author{R.~Avakian}  \groupyerevan
\author{A.~Avetissian}  \groupyerevan
% ...

\collaboration{The HERMES Collaboration} \noaffiliation